# A DETAILLED WEATHER DATA GENERATOR FOR BUILDINGS SIMULATION

L. ADELARD*, H. BOYER, F. GARDE, J-C GATINA
*Université de la Réunion, Laboratoire de Génie Industriel, Equipe Génie Civil*
*15 Avenue René Cassin, 97705 Saint Denis Messag Cedex 9, Ile de La Réunion, France*
*Email : adelard@iremia.univ-reunion.fr*

*Abstract*

Thermal buildings simulation softwares need meteorological files in thermal comfort, energetic evaluation studies. Few tools can make significant meteorological data available such as generated typical year, representative days, or artificial meteorological database. This paper deals about the presentation of a new software, RUNEOLE, used to provide weather data in buildings applications with a method adapted to all kind of climates. RUNEOLE associates three modules of description, modelling and generation of weather data. The statistical description of an existing meteorological database makes typical representative days available and leads to the creation of model libraries. The generation module leads to the generation of non existing sequences. This software tends to be usable for the searchers and designers, by means of interactivity, facilitated use and easy communication. The conceptual basis of this tool will be exposed and we'll propose two examples of applications in building physics for tropical humid climates.

**1. Introduction**

**2. Conceptual basis**

**3. Description of RUNEOLE**

 *3.1 Introduction*

 *3.2 Module of climate characterisation.*

 *3.3 Module of weather data generation*

 *3.4 Physical modelling module*

**4. Applications in building physics**

 *4.1 Description of the code CODYRUN.*

 *4.2 Presentation of the buildings.*

 *4.3 Method to provide meteorological data.*

 *4.4 Results in HVAC energy needs study.*

 *4.5 Results for the prototype thermal comfort.*

**5. Conclusion**

______
* Corresponding author
Tel : +33 2 62 96 28 96 Fax : +33 2 62 96 28 79
E-mail : adelard@iremia.univ-reunion.fr

# A DETAILLED WEATHER DATA GENERATOR FOR BUILDINGS SIMULATION

L. Adelard*, H. Boyer, F. Garde, J-C Gatina

Laboratoire de Génie Industriel, Equipe Génie Civil

Université de la Réunion

15, Avenue René Cassin

97715 Saint Denis Messag Cedex 9

*Abstract*

Thermal buildings simulation softwares need meteorological files in thermal comfort, energetic evaluation studies. Few tools can make significant meteorological data available such as generated typical year, representative days, or artificial meteorological database. This paper deals about the presentation of a new software, RUNEOLE, used to provide weather data in buildings applications with a method adapted to all kind of climates. RUNEOLE associates three modules of description, modelling and generation of weather data. The statistical description of an existing meteorological database makes typical representative days available and leads to the creation of model libraries. The generation module allows the generation of non existing sequences. This software tends to be usable for the searchers and designers, by means of interactivity, facilitated use and easy communication. The conceptual basis of this tool will be exposed and we'll propose two examples of applications in building physics for tropical humid climates.

## 1. Introduction

Coherent meteorological data are always needed in system sizing or predicting energy needs in buildings. In the last decade, hundred simulation codes were elaborated, but only about ten weather data generators were established. Few types of meteorological data can be provided, such as typical days extracting from an existing database, or artificial data generated by a software using a serie of pre-established models, when no hourly database is available on the site.

In building physics, Hui [1] as Lam [2] exposed that average data (typical meteorological year, artificial one-year data generated) could be used in long term consumption evaluation. Extreme data are used in sizing studies for example for HVAC system. In each kind of climate (temperate, tropical), for each kind of buildings (individual or collective, heavy or light structure), climatic factors don't have the similar influence. So, in the design phase, it is important to have current and extreme conditions for each climatic factor. This follows what says Lam when he explains that you don't need only one typical meteorological year, but one for each climatic factor [2]. The table 1 gives a summary of each type of meteorological data applications.

*Table 1 : Classification of existing means to provide meteorological database*

The conclusion of such an analysis is that a complete tool to provide coherent climatic data must be able to propose :

---

* Corresponding author
Tel : +33 2 62 96 28 96    Fax : +33 2 62 96 28 79
E-mail : adelard@iremia.univ-reunion.fr

- Typical days representative of current, average, and extreme weather conditions for a site. We'll use the thought process of Le Chapelier and Sacre [3-4]
- Mathematical or physical models to generate a sequence needed by the user if it is not to available in the existing database. Models must also be available if a climatic variable is not available for an existing sequence. As models are dependant of a type of climate, the software will have to elaborate several libraries including parameters, mean relative error for each proposed model to the user. The algorithm used to generate the data follows the wishes of the user.
- Complete data generation if no database at all is available for a site.

The software presented in this paper includes these three purposes. We'll expose its application in two cases of study. The first one will be an energy needs evaluation in a collective building. The second is a comfort study in a prototype of individual house. It'll be able then to explore a new way of study by designing a building envelope and testing it in a very large climatic solicitations configurations.

**2. Conceptual basis**

Methods to dispose of climatic data appear with the evolution of computerised methods for determining man's artificial environment. First, simplified methods as "bin data" for temperature [5], gave estimations of heating or cooling loads. The concept of typical meteorological years was created to provide hourly data representative of an average year on the site [6] for more accurate estimation of the energy needs. Selected meteorological data are identified by the closeness of the cumulative distribution functions. TMY became then the reference for testing all the methods used to provide meteorological data. As a lot of authors said that for a specified building [6,2], it is often rather difficult to determine which type of weather will be most critical or typical. That implies that for various kind of architecture, several TMY are needed (it is necessary to vary the influence of each climatic variables in the choice of the representative year).

"Representative days" have been created to reduce the computational time. It can be used in simplified simulation and design tools. These days have to represent the typical climatic conditions. They correspond to average, current, and extreme conditions for each climate element with informations on the frequencies of each conditions. They can be chosen with the help of a statistical classification [3], following the steps proposed by Le Chapelier. First, an application of the principal components analysis (PCA) is applied to each climatic variable to determine the best factors describing each variable. Then, an automatic ascending classification is applied to these factors to find the best classes of days.

Data can also be generated by a weather data generator. This kind of code uses statistical analysis to model the climatic variables. Generally, entries of such tools are monthly mean daily data. So each climatic variable is generated by random choice using its statistical distribution (daily and hourly), autocorrelation, and, when possible, taking into account its interactions with the other variables.

Degelman [7-8] created a generator which used the monthly mean data as entries of the code, so it can be used in all sites where monthly data are available. However, this code is not suitable for the tropical humid climates. In fact, interactions between variables are not really taken into account in this software. Knight [9] included a similar algorithm in TRNSYS (only few modifications were brought). It gives an average year, corresponding to the main climatic statistics of the site. For Van Paassen and Dejong [10] the adopted approach was to separate all climatic variables in two parts. The first part is dependent of the solar radiation and the second part is generated independently with statistical means (random part). Models used are obtained by signal processing, correlation functions, and stochastics models. Data are then generated taking into account of its frequency in the year. Thianzhen Hong [11] proposed the use of stochastic models with external variables to take into account of the interactions between solar radiation and temperature. Sacre [12] used probability matrices and correlation functions to generate only artificial representative days.

The generators are usually used to provide data on a whole year, but recently Degelman [13] proposed to use just one week instead of a whole month. Van Paassen and Sacre generate only typical sequences. This could lead to the same results and limiting the computational time. Weather data generators are generally used when no database is available on the site. However, both tropical and temperate models must be available for a wide using.

To conclude, weather data generator are usually defined by the climatic variables used as entries, simulated variables, and type of models integrated. The table 2 exposes these different points for each generator found in the bibliography.

*Table 2 : Characteristics of the different weather climate data generators.*

RUNEOLE was elaborated considering the following points :
- No software actually can be used to describe an hourly meteorological database and give a statistical description of each climatic variable existing in it. Statistical description goes from a simple statistic table where means and daily maxima for a specific period (month, season) can be done to the elaboration of models for artificial data generation, or another need of the user. Statistical description also includes representative days searching.
- The generator must be able to furnish non existing climatic variables in an existing database.
- Instead of generating one year of data, we want to generate climatic sequences, as asked by the user considering its specific needs. This implies that the entries of the climate generator will be able to be modified by the user. If he wants for example a warm sequence for the humid season, our generator must be able to provide the result and to specify if necessary the frequency of this sequence, and the climatic conditions in which this may occur. Our purpose is to generate not only traditional climatic sequences, but also to be able to simulate non current situations.
- The software must be used in several kind of climates. This leads to the necessity of a multimodel approach.

The algorithms created for the previous climate generator will be used to create RUNEOLE, with some modifications. Models used in RUNEOLE will be structured by level of complexity. First, there will be correlation functions determined with regressions. Bibliography is rich of correlation functions between solar data such as global and diffuse solar radiation, insolation, nebulosity. As in mentioned generators, models are separated by daily and hourly models. The table 3 mentions all the empirical models used at this level. The generated variable is in bold character, and the predictors are in simple character.

*Table 3: Correlation functions integrated in RUNEOLE.*

The second level is made with stochastic models. Bibliography exists for the elaboration of AR or ARMA models. For the theory, we can quote Box and Jenkins [14]. For the applications, we can quote Sbai [17], or Nfaoui [16] for the wind speed generation and Graham [17] for the clearness index generation. This kind of model will be used for generating the first variable of the artificial sequence. Using stochastic model includes a study of distribution laws for each climatic variable. We used Weibull's law for wind speed [15], Saulnier's law [18] for clearness index in tropical climates, and Liu and Jordan's law for temperate climates [19].

The third level is for the taking into account of the interactions between parameters such as temperature, relative humidity, wind speed and solar radiation. Using spectral analysis, we saw that, at the hourly step, interactions were non linear, so we decided to use black box models, with neural network models. It has also been used for the sky temperature determinations [20]. Autoregressive models (also used by Thianzhen Hong) are actually elaborated.

We have to remind here that the sequences will correspond to the wishes of the user. The structure of such a generation will be described in the description of the code. We exposed here only the type of models involved in the elaboration of RUNEOLE. The mathematical formulations of the main models will be exposed in a future paper.

## 3. Description of RUNEOLE

*3.1 Introduction*

RUNEOLE combines the methods of characterisation, modelisation, and generation of meteorological datas, leading to a generalist tool, obtaining the most informations from a database. It is possible then possible to use existing database to create necessary models to extrapolate climatic sequences for other sites where no data is available. RUNEOLE is made for designers and searchers. Use by designers involves an easy use resulting in the software's rapidity in furnishing wished outputs and first exploitations of the results with analysis commentary. For this, we'll use evolutive analysis window where users can store their own commentaries. Extracted or generated files will have to respect the format needed by other codes.

RUNEOLE is elaborated on a PC computer often used in design departments. However, extracted sequences can be exported on other platforms such as Macintosh or UNIX stations. The programmes has been made on Matlab to dispose of the multiple toolboxes. However, it will be traduced in C language in a further work. The description of each module will let us show some navigation windows of this software. The figure 1 summaries the general objectives and imperatives of RUNEOLE.

*Figure 1 : Description of RUNEOLE imperatives*

*3.2 Module of climate characterisation.*

Basic and complex statistics analysis are found in this module. Representative days and particular sequences are also searched in this part. Basic analysis describes the database by showing statistical tables and histograms for each climatic variables (figure 2). For this, the climatic variables are represented by a statistical indicators adapted to the users needs (daily mean, daily maxima, daily amplitude), and histograms are given for this indicators for the wished period. The "univariable" part is made for analysing each variable separately. The "multivariable" part provides a contingency table, and leads to the Chi-2 test for the dependence study.

*Figure 2 : Basis statistical analysis result windows (Histogram for the humid season for global solar radiation).*

Complex analysis includes each modelisation window (distribution laws, correlation function, stochastic and neural network modelisation (the two windows are presented at figure 3)). Here, the period, the variable (hourly or daily data) are chosen, and the kind of model or law used for the study.

*Figure 3 : Windows of complex statistical analysis*

This part can be used to elaborate models for other applications (for example, for integration in TRNSYS types). Searching sequences can be made with the users criteria. RUNEOLE chooses automatically the methods according to the entries of the users criteria. If the user just wants to choose the variables he wants to take into account, the code uses automatic classification. If the user goes on and choose numerical criteria (for example, mean temperature between 30-32°C), there will be just a selection of data. The figure 4 presents two windows used in selecting the choices of the user and proposing the selected sequence to other codes.

*Figure 4 : Representation of windows of climatic sequences determination*

*3.3 Module of weather data generation*

The second module is devoted to the generation of artificial sequences. First, the user sets the wanted conditions (for example, a clearness index of 0.75 and a wind speed of 3 $m.s^{-1}$), the variables to be generated, and the number of days. RUNEOLE generates first data for global solar radiation and wind speed. At each step, the software finds automatically what variable can be generated, and which model can be used; then, the definite choice will be let to the user. So, if insolation, beam, or diffuse radiation are needed, correlations are proposed to the user. Then temperature and humidity are generated with the selected model (actually, the best model is neural network). Such a process makes the data coherent because the software generates a new variable according to the already generated data. If a model is needed for the generation with no available parameter, RUNEOLE goes automatically in the complex statistical analysis module and determine the parameters. The figure 5 shows the windows for the choices of conditions and models for a generated sequence.

*Figure 5 : Generation of artificial climatic sequences*

*3.4 Physical modelling module*

The third module leads to a physical modelisation. It leads to sky temperature and thermodynamic variables determination such as vapor pressure, enthalpy, dew point temperature. Secondly, spatial extrapolation of database will be introduced taking into account the altitude difference, the longitude and latitude of the new site.

Here, we have exposed the capacities of RUNEOLE. In the next part, we are going to illustrate the use of this code in building physics applications. The first is relative to an evaluation of cooling loads in a tropical humid climate using representative or generated sequences. The second is a thermal comfort study in a prototype house.

## 4. Applications in building physics

Reunion island is situated by +21° South Latitude and 55° East longitude, next to Madagascar island. Situated in a tropical zone, the year is theoretically divided in two seasons:

- In the humid season (from November to April), the island is close to the inter tropical convergence zone and cyclones perturbations may occur.
- The fresh season is from May to October, when the climate of the island is influenced by trade winds. The island is highly mountainous. There are smooth bents on the cost zones, which increase quickly toward the centre of the island. The centre is made of three cirques which give a very contrasted relief. Such complexity on a small surfaces (2500 km²), gives a lot of micro climates.

Demography, associated with the importation of unadapted construction models leads to the development of buildings totally unadapted to humid climates. The sensations of thermal discomfort in these buildings have increased the energy consumption by the use of individual air conditioning systems. It was then necessary to promote bioclimatic or energy efficiency architecture [21]. This development needs precise informations about the effect of climate patterns on buildings.

For energy applications in buildings, meteorological data are needed for simulations. In Reunion Island, the building thermal behaviour depends strongly on external climatic conditions. Heat transfers are due to convection, conduction and above all to solar radiation. That is different from the temperate climate zones. Wind speed and air humidity are also important in studies of the human comfort in buildings [22].

*4.1 Description of the code CODYRUN.*

The HVAC system being studied will be simulated with an airflow and thermal buildings simulation software, CODYRUN. A description of the model can be found in [23] and [24]. This multizone software has been designed to suit the needs of researchers and professionals. It was developed on a PC micro computer with Microsoft windows user-friendly interface. CODYRUN integrates both natural ventilation and moisture transfers. It offers to the user a wide range of choices between different heat transfer models. The software is close to ESP [25] for the thermal aspect. The airflow model [26] can be compared to Airnet. The choice of models will depend on the objective of the simulation:
- Study of temperatures regarding to the choice of components of the buildings.
- Estimation of the daily energy consumption.

The software is divided in three parts:
- A module of description of the building by means of zones (from a thermal point of view), inter-zones (zones separations), components (walls, glass partitions, HVAC systems, ...)
- In the module of simulation, Boyer [23] used the systemic analysis to build mathematical model. So the building is decomposed in few zones. The technique of nodal discretisation of the space variables by finite difference is then used. The sets of equations for the zone thermal model [27] can be condensed in a linear steady state evolution equation:

$$\mathbf{C}\frac{dT}{dt} = \mathbf{A}\,T + \mathbf{B}$$

For the study of HVAC systems, the code includes three types of models. These models have been elaborated from a very simplified one to a very detailed one. The first model is just supposed to supply the demand of cooling loads with an ideal control loop (no delay between the solicitations and the time response of the system). The available outputs are initially the hourly cooling consumption without integrating the real characteristics of the HVAC systems. The second and the third model are more detailed that the first one. The time step was reduced to one minute in order to take into account the on-off cycling and the control of the systems.
- In the third module of exploitation of results, the outputs are plotted and exported to classical spreadsheets to be exploited.

*4.2 Presentation of the buildings.*

The first studied case is a residential building. After consulting the local low cost housing institutions and the new housing statistics for Reunion Island, we found that the most common dwelling found is the T3/4, consisting of two bedrooms, a living room and a veranda (figure 6).

*Figure 6: Description of the studied building*

We'll analyse here the HVAC energy needs for a flat situated on the top of the building, beneath the roof. The bedroom is oriented towards the East, and the living-room is oriented towards the West. We'll consider the flat as a two zones building (living-room zone and bedroom zone). The dwelling will be internally solicited by loads constituted by four inhabitants (two adults and children), and by electric lamps in each room (table 4).

*Table 4: Internal loads for the collective dwelling*

*Table 5 : Thermophysic description of the materials used in the studied building*

In table 5, the thermophysic characteristics of the two studied case are presented. Simulation and experimental measurements showed that the high internal temperature (33°C) in a closed dwelling made necessary the installation of HVAC system used from 8 p.m. to 6 a.m. (figure 7).

*Figure 7 : External and internal temperature for the closed dwelling*

The second studied case is an individual prototype house made with aluminium panels-. Its surface is about 65 m², with a veranda, one bedroom, and a living-room. False ceiling is made of glass wool. Openings are constituted with a 1 m² window and a sliding picture window. The house casing is made of a metallic frame and each wall is made of an assembly of sheet steel, glass wool and plasterboard (figure 8).

*Figure 8 : Prototype house*

The measurements experimental campaign showed that thermal comfort is obtained dry season (figure 9). For the thermal comfort study, few methods are proposed, taking into account of the human sensations [28-29]. ASHRAE Standard 55 proposed a summer comfort zone function of temperature (from 22.8°C to 26.1°C) and relative humidity levels for health considerations (60% max.), but theses margins can be modified for tropical climates [28]. Berglund and Cain [30] considered that relative humidities above 65% were associated with unacceptable air quality judgements. In this study, we used the Givoni's zones for the determination of thermal comfort conditions. These two zones placed on a psychrometric chart are relative to satisfying conditions in no wind case (zone 1) and satisfying conditions for a wind speed inferior to 1.5m.s$^{-1}$ (zone 2) [31].

*Figure 9: Comfort diagram for the dry season for the prototype house (experimental data).*

The simulations we will run will aim at evaluating thermal comfort in humid season conditions, using specific climatic sequences to show the relationship between outdoor weather parameters and the indoor conditions. A first interesting approach was made by Givoni [32], analysing the relation between outdoor and indoor air temperature for closed dwellings. Givoni used non identified weather conditions, and several building configurations to study the influence of mass and night ventilation. Here, we'll make simulations for a given building, with different climatic configurations to study the influence of different climatic parameter.

*4.3 Method to provide meteorological data.*

The following database of Gillot meteorological station including the main variables : dry air temperature, relative humidity, wind speed and direction, nebulosity, insolation, diffuse and global solar radiation as hourly data. The variables taken into account in this study are air temperature, relative humidity, wind speed and solar radiation. In our analysis by histograms, the mean radiation is found to be of 5900 Wh.m$^{-2}$.day$^{-1}$, and the maxima is of 8740 Wh.m$^{-2}$.day$^{-1}$ (figure 2). The mean daily temperature is of 27°C, with a daily maxima of 35°C. The mean wind speed is 5 m.s$^{-1}$. Study of representative days for each of these variables shows the main hourly evolutions. Here, we present the three evolutions given by RUNEOLE. The first is relative to low radiation (about 1000 Wh.m².day$^{-1}$). The second evolution is relative to mean radiation mainly received in the beginning of the day from 6 to 14 hour to high radiation, and the third is relative to great sunny days (figure 10).

*Figure 10 : Three representative evolutions for the solar global radiation for the humid season*

An automatic classification using the four variables has lead to the selection of the current sequences A, B, C, D, E. The extreme sequences are E for the extreme humidity and F for the extreme temperature. All of the sequences (expect sequence E) include high global solar radiation. The table 6 sums up the climatic sequences chosen for the study, and their frequencies in humid climate. The following parts show the different results.

*Table 6: Description of the sequences.*

*4.4 Results in HVAC energy needs study.*

Haberl and al. [33] showed that it was useful in evaluating energy use to use both measured data and typical meteorological year. The mean data of typical meteorological year was leading to a good sensible demand estimation, and the measured data allowed a better latent demand estimation. In our study, several weather sequences of the humid season will be used to simulate HVAC behaviour by the following ways:

- The first way consists in simulating the building by using the three most humid and hot month in the year which are December, January, and February (120 days). We calculate then, as a reference, the mean daily energy consumption for cooling for these months (table 7).

*Table 7. Cooling capacities for the different sequences.*

- The second way consists in simulating the building by using the weather sequences to analyse precisely the energy consumption in each of these situations. We have carried out an energy needs study for each of the sequences presented in table 6. Each of these sequences have about 5 days, because the weather varies relatively quickly in the humid season. For each of these sequence, we give the mean and the maximum energy needed per day. We separate the energy need for dehumidification (the latent cooling capacity) and the energy need for lowering the temperature (the sensible cooling capacity). The mean daily cooling capacities (MEAN), and the maximum daily cooling capacities (MIN) are presented in the table 7.

The results bring to the light the influence of coupling conditions of wind, temperature, humidity, and radiation. The sequence A shows the importance of coupling wind and temperature. This situation requires an important sensible capacity. Latent cooling is also important in this sequence, but it is not the worst situation for the latent cooling needs. The maximum sensible cooling capacity is found for this case. The sequence B includes similar characteristics but with a lower mean temperature. In this case, energy needs are equivalent to the mean results presented in the first lines of table 7.

The sequence C contains more hot days with high radiation, but with a medium wind. Even with great convective heat transfer due to the medium wind speed, the sensible cooling capacity is above the mean results. However, it is not the most important demand. Sequences with high humidity are frequent in humid season. Latent consumption doubles in these cases. It is the case of the sequence E.

For the sequence F, we found that for dry conditions, the most important sensible cooling capacity (22 kWh) for a maximal temperature of 35°C for the day. We found the same result as for sequence A, but for a lower maximal temperature (30° C). It must be caused by the difference of wind speed between these two sequences. Effectively, sequence A is less windy as sequence F.

We showed the sensitivities of the building to the climatic factors. In the studied case, the main parameters is global radiation and temperature. It is quite difficult to show clearly the influence of these two factors separately. We noticed that for 1.5°C above mean temperature, energy needs rise of 6 kWh, for similar wind speed. Influence of humidity is not negligible rising the latent demand from 2 to 6 kwh for a relative humidity getting from 60% to 90%. In 40% of time, sensible consumption is above the mean calculated mean (figure 11) and in 25% of time the sensible consumption is above 15 kWh.

*Figure 11 : Distribution of latent and sensible energy needs for the humid season*

Finally, the advantage of using such sequences is to determine the main characteristic of the HVAC power demand rapidly, with knowledge on building sensitivity to the different climatic factors.

*4.5 Results for the prototype thermal comfort.*

As said earlier, the dry season leads to thermal comfort conditions in the prototype house. In humid season, in rainy conditions, thermal comfort is verified with an indoor air speed of 1.5 m.s$^{-1}$. In case of low wind speed in the house, the high relative humidity will lead to the discomfort situation (figure 12).

*Figure 12 : Thermal comfort diagram for the sequence F*

Worst conditions for thermal comfort are found for high temperatures. Effectively, for the sequence C, 50% of points are outside of Givoni's zone 2, and 40% for the sequence A. We noticed that we use here the zone 2, relative to an air speed of 1.5 m.s-1. As the air speed are low in the closed house, we can say that this type of sequence leads automatically to discomfort situations.

*Figure 13 : Thermal comfort diagram for the sequences A and C*

*Figure 14 : Comparison of internal temperature in the house for the three sequences A, B and C*

We present at figure 14 the observations made for the sequences A, B, and C. In these simulations, we created a little opening in the house's attic. We can see then the progressive rising of internal temperature from the sequence A to the sequence C. Conditions for the sequence a and B are similar. Sequence A and B lead to satisfying comfort conditions, but sequence C leads again to discomfort situations.

In this study, we showed how to understand quite rapidly the interactions between climatic factors and indoor thermal conditions using characterised sequences. A further step should model the relation between indoor air temperature and outdoor conditions. Givoni [31] showed that the best correlation exist between the outdoor average and the indoor maximum. Buildings characteristics as heavy or light structure will have to be considered.

**5. Conclusion**

The main objective of this paper was to illustrate the use of the new code RUNEOLE to provide coherent and weather sequences for building physics applications. The weather sequences were chose according to characteristic (current or extreme situations for climatic factors) and criteria based on professional needs. Specific study of the cooling capacity were made for the collective dwelling. As we have no reference year for the island site, we just showed that the choice of the sequence could lead to very various results. This study allows us to confirm a previously known result, the worst conditions appear for mean temperature, high radiation and above all, breeze and high relative humidity. The simulations also showed the great latent consumption for the rainy sequences.

So, we showed that the use of such sequences could give faster the mean and the maxima needs for HVAC system. But this can also show in which conditions occur each kind of consumption.

The main sequences for the humid sequences have also been used for thermal comfort determination for a prototype house. We showed that the main climatic factor was solar radiation and external temperature. A ventilation is needed for a satisfying thermal comfort.

This kind of simulation is useful to trained the buildings under extreme climatic conditions. The next stage of this study should be to compare our results with this obtained by using a typical meteorological year, following the steps adopted by Haberl. A further study, leading to a more detailed of climatic influences on different kind of buildings is envisaged.

A validation of the code RUNEOLE must be made. Two ways will be followed for this : a statistical validation for all the models integrated, and an indirect validation using the results of simulation with test reference year and RUNEOLE sequences to verify energy evaluations [2]. This validation will also integrate the test for different kind of climates.

ACKNOWLEDGMENTS

This research is supported by a grant from the Conseil Regional of Reunion Island. The weather data were supplied by the Meteo France services in Reunion Island.

| Designation | Utilisation | Advantages and Inconvenients |
|---|---|---|
| « Bin » Data Degree-days | Heating consumption evaluation | Insufficient information volume Easy use (table) |
| Multiple years (MY) | Energetic consumption in building and thermal evaluation | Best precision Great computing time Big volume of data |
| Typical meteorological year (TMY) Test reference years (TRY) | Idem | Good precision of the mean energy needs Possible choice of a non adapted year considering the building sensitivities |
| Representative days and sequences Short reference years | Solar and HVAC systems sizing | Time gain Possible over or underestimation of systems sizing |
| Weather data generators | Supplying non existing data Systems sizing, energy needs evaluation | Difficulties to model climatic variables |

**Table 1: Classification of existing means to provide meteorological database**

| Authors | | Degelman, [1976, 1991] | Van Paassen [1979, 1984] | Sacré [1984] | Thianzhen Hong [1996] |
|---|---|---|---|---|---|
| Type of Models | Empiric relations | × | × | × | |
| | Stochastic models (autoregressive) | × | × | × | |
| | Stochastic models with external variables | | | | |
| data needed as input | | | Monthly means | Monthly means | Representatives days, probability matrices of transition | Monthly means |
| Generated Data | Hourly | × | × | × | × Temperature and global solar radiation |
| | Daily | × | × | × | × Temperature max. and min. et amplitude global solar radiation |
| Models assembly | | | Generation of sky conditions, then, of the other variables | Generation of global solar data, then , of the other variables | | Simultaneous generation |

**Table 2 : Characteristics of the different weather climate data generators.**

| Type | Désignation (Authors) |
|---|---|
| Insolation - **Clearness index** | Angström Black |
| | Hay |
| Clearness Index - **insolation** | Angström Black Inverse |
| | Hay inverse |
| Global radiation - Insolation - **Diffuse radiation** | Klein |
| | Page |
| Insolation - Global radiation - **Diffuse radiation** | Gopinathan 1 |
| Insolation - **diffuse radiation** | Icqbal |
| Clearness index - Global radiation-beam radiation | Erbs |
| Clearness index - beam radiation - Nébulosity | Castagnoli |
| Clearness index - **Nébulosity** | Barri |
| Insolation - **Nébulosity** | Rangarajan |
| Global radiation- Insolation - Clearness index -**Diffuse radiation** | Gopinathan2 |
| Insolation - Global radiation- Clearness index - **Diffuse radiation** | Soler |

**Table 3: Correlation functions integrated in RUNEOLE.**

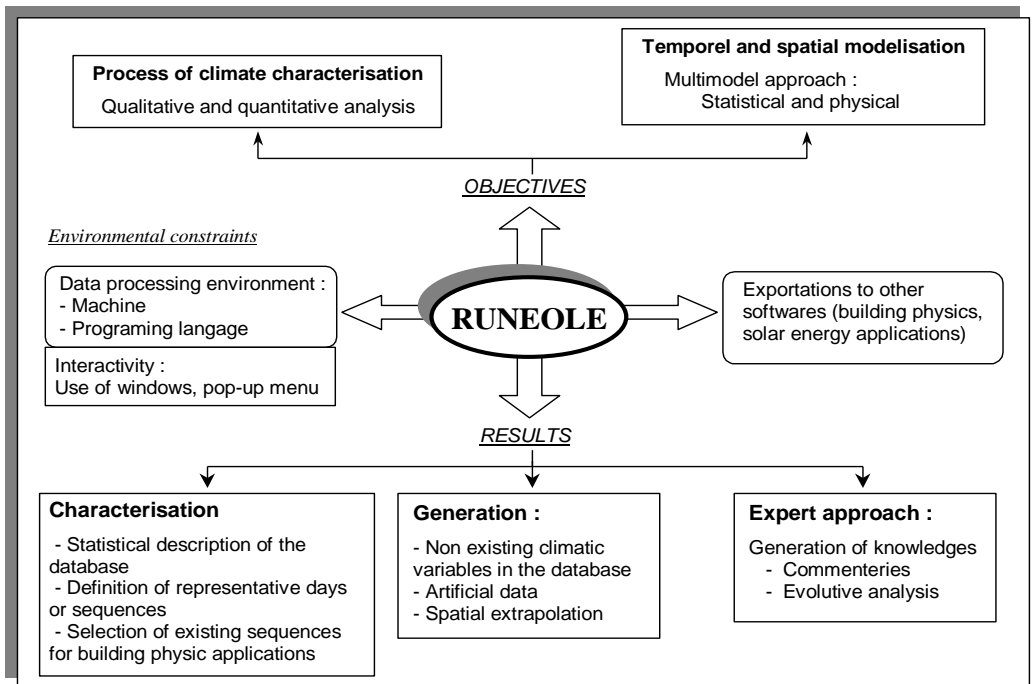

**Figure 1 : Description of RUNEOLE imperatives**

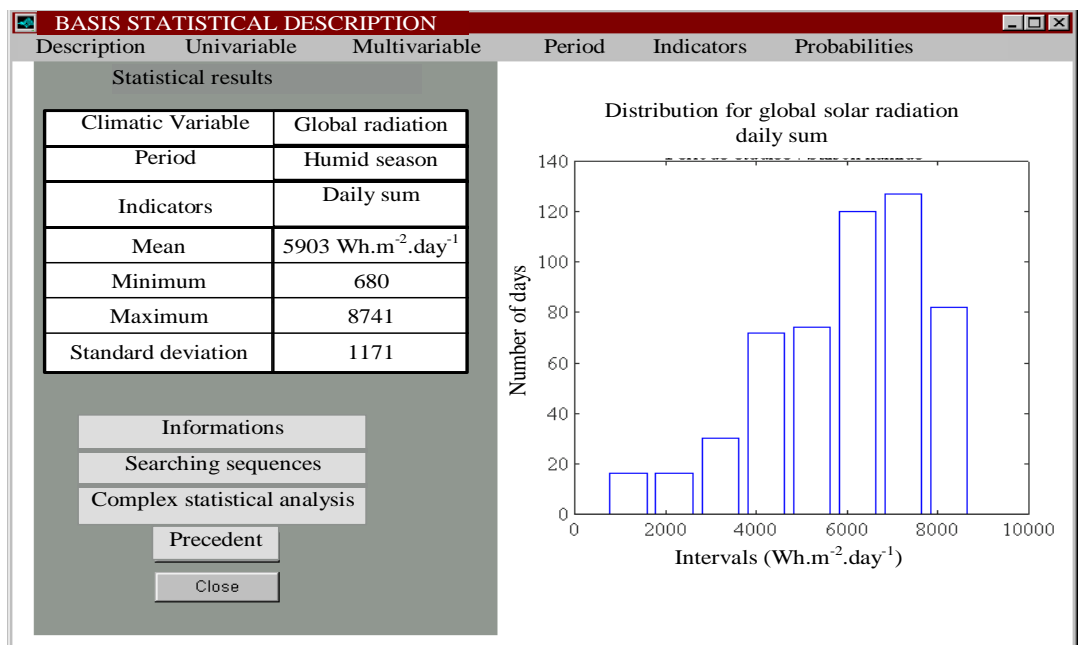

**Figure 2 : Basis statistical analysis result windows (Histogram for the humid season for global solar radiation).**

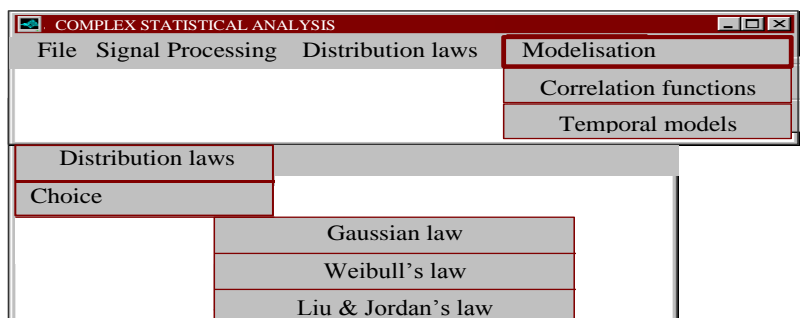

**Figure 3 : Windows of complex statistical analysis**

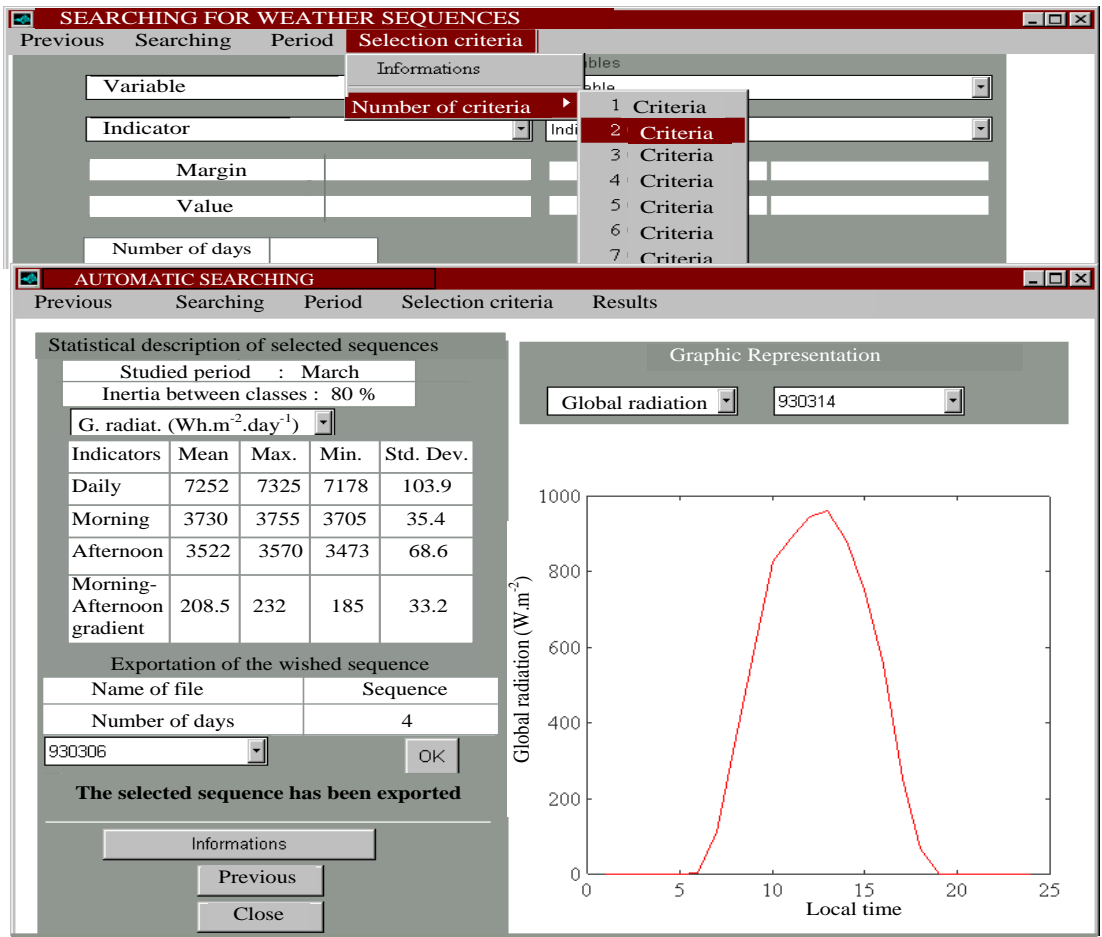

**Figure 4 : Representation of windows of climatic sequences determination**

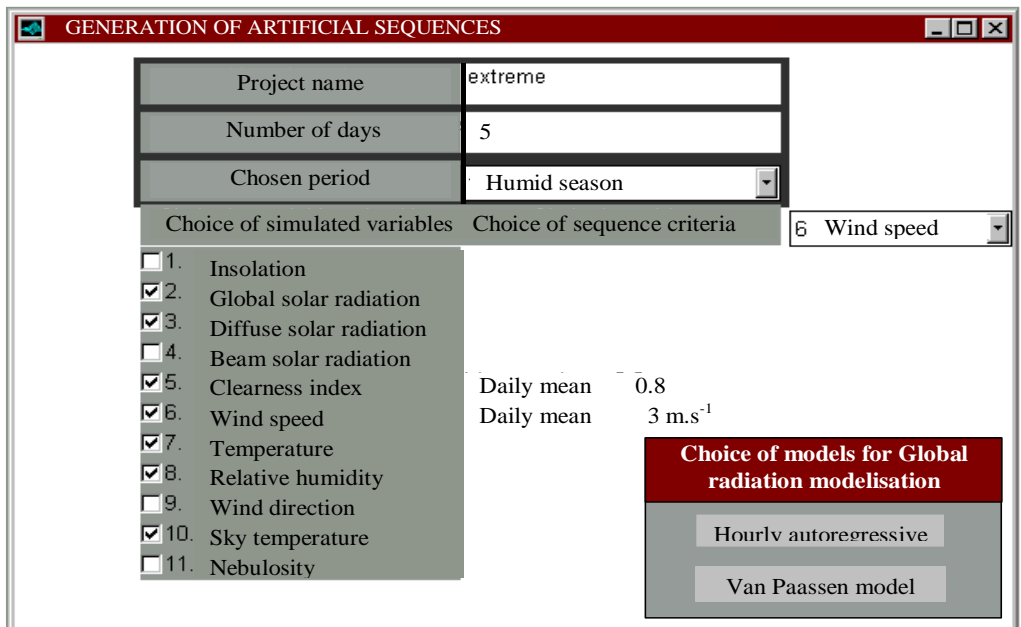

**Figure 5 : Generation of artificial climatic sequences**

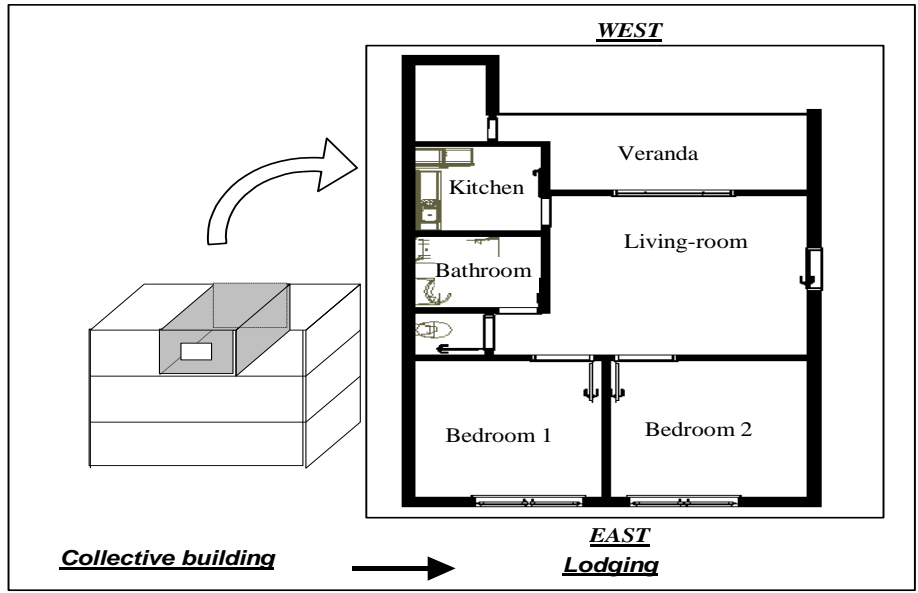

**Figure 6: Description of the studied building**

| Internal loads | Sensible load (w) | Latent load |
|---|---|---|
| Adult | 60 | 60 |
| Child | 40 | 40 |
| Lightning bedroom 1 | 100 | 0 |
| Lightning bedroom 1 | 100 | 0 |
| Lightning living room | 300 | 0 |

**Table 4: Internal loads for the collective dwelling**

| Colective dwelling | Materials | Thermal resistance (m².W$^{-1}$) | External absorptivity $\alpha$ |
|---|---|---|---|
| Roof | Concrete 16 cm | 0.1 | 0.7 |
| External wall | Béton banché 20 cm | 0.11 | 0.7 |
| Internal wall | Placoplâtre + lame d'air 5 cm | 0.2 | - |
| *Individual house* | | | |
| External wall | Steel sheets | 40 | 0.4 |
| | Glass wool | 0.04 | 0.6 |
| Internal wall | Plaque de parement de placoplâtre | 0.42 | |

**Table 5 : Thermophysic description of the materials used in the studied building**

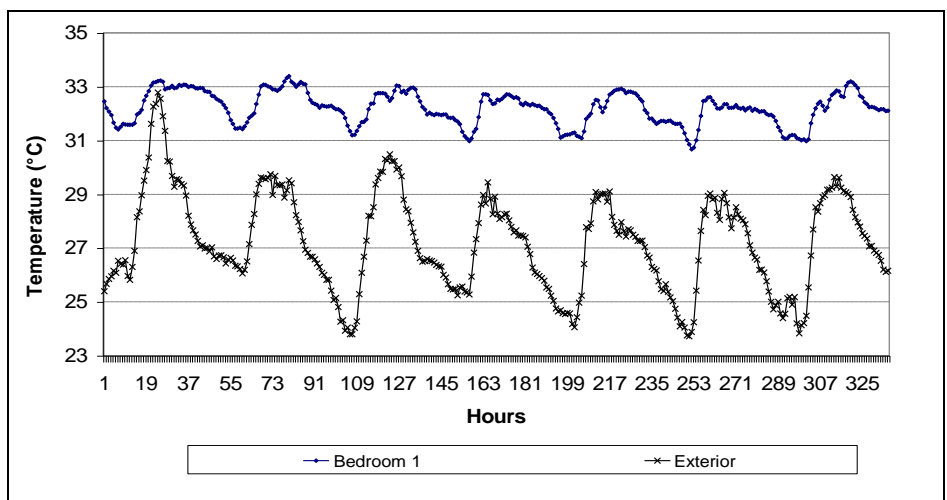

**Figure 7 : External and internal temperature for the closed dwelling**

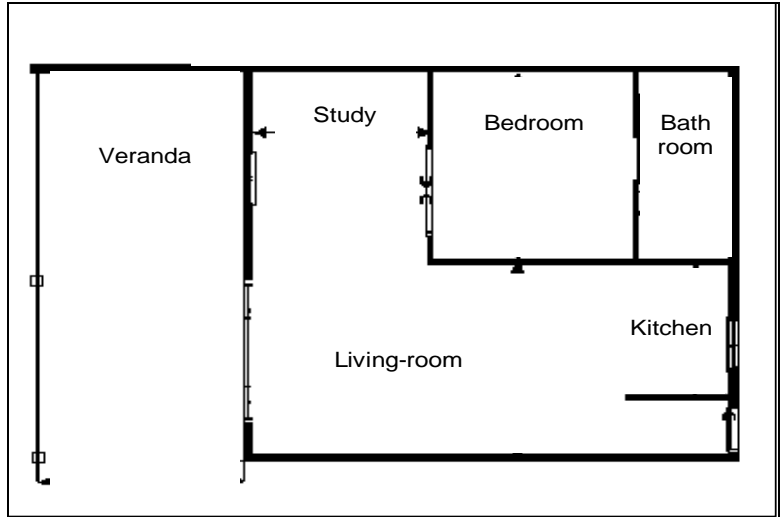

**Figure 8 : Prototype house**

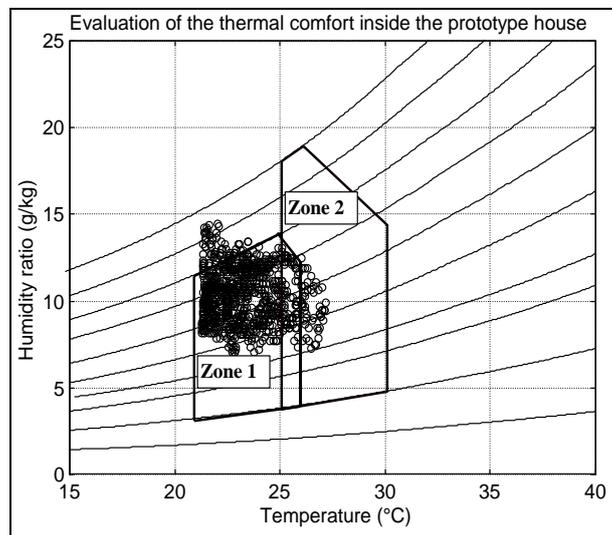

**Figure 9: Comfort diagram for the dry season for the prototype house (experimental data).**

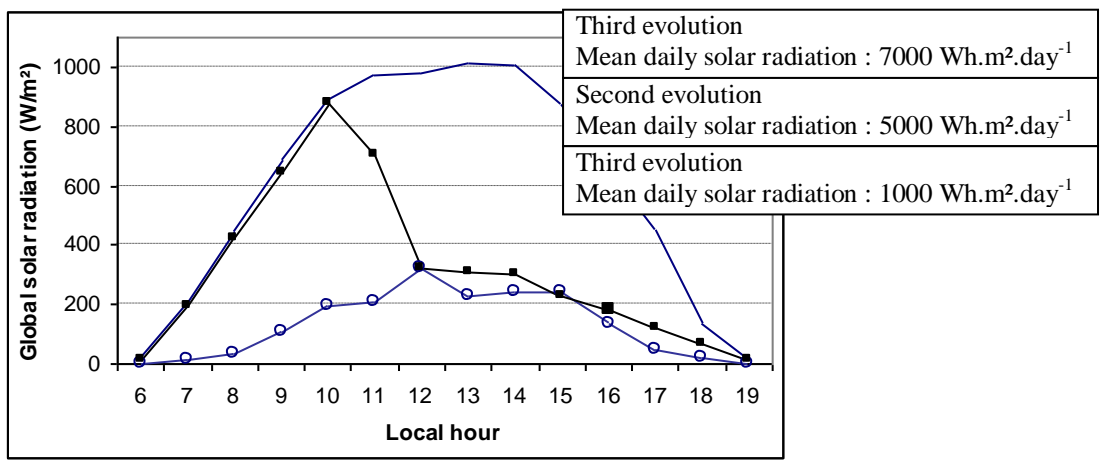

**Table 6: Description of the sequences.**

| Sequences | Frequency | Description |
|---|---|---|
| Sequence A | 20 % | High to very high radiation (from 5700 to 8400 Wh.m-2)<br>Breeze (0 to 3 ms-1)<br>Mean temperature:27°C     Maxi temperature:30.7°C |
| Sequence B | 10 % | High to very high radiation<br>Breeze<br>Mean temperature:25.5°C    Maxi temperature:30.8°C |
| Sequence C | 15 % | High to very high radiation<br>Medium to strong wind (from 3 to 9 m.s$^{-1}$)<br>Mean temperature: 28°C    Maxi temperature:33.2°C<br>*(This sequence is current when a cyclonic perturbation is in proximity of the island)* |
| Sequence D | 16 % | Very high radiation with medium to strong wind<br>Mean temperature: 26°C     Maxi temperature: 32 °C |
| Sequence E | 15 % | Sequence of rainy days. *(This kind of situations is relatively frequent in the wet season).*<br>Low radiation<br>Mean temperature: 26°C     Maxi temperature: 30°C |
| Sequence F | 5 % | Dry and hot sequence. *(This kind of sequence is relatively rare in the wet season, however, we used it because it occurs a maximal temperature of 35°C).*<br>High radiation<br>Mean temperature: 27°C     Medium wind  (3 to 6 m.s-1) |

**Table 7. Cooling capacities for the different sequences.**

| Type of sequence | Nature of the daily capacity | Sensible cooling capacity (kwh) | Latent cooling capacity (kwh) | Total cooling capacity (kwh) |
|---|---|---|---|---|
| Mean Conditions | MEAN | 10 | 2.6 | 12.6 |
|  | MAX | 18.8 | 5.7 | 21.5 |
| Sequence A | MEAN | 16 | 2.6 | 21.2 |
|  | MAX | 18.6 | 3.5 | 24.5 |
| Sequence B | MEAN | 15.6 | 1.8 | 12.3 |
|  | MAX | 16.4 | 2.8 | 14.3 |
| Sequence C | MEAN | 16.2 | 2.5 | 21 |
|  | MAX | 17.5 | 3 | 22.8 |
| Sequence D | MEAN | 13.5 | 2.5 | 16 |
|  | MAX | 18 | 4.7 | 22.7 |
| Sequence E | MEAN | 8.5 | 4 | 12.5 |
|  | MAX | 14 | 5.7 | 18 |
| Sequence F | MEAN | 17 | 2.8 | 19.6 |
|  | MAX | 18.6 | 3.4 | 21.5 |

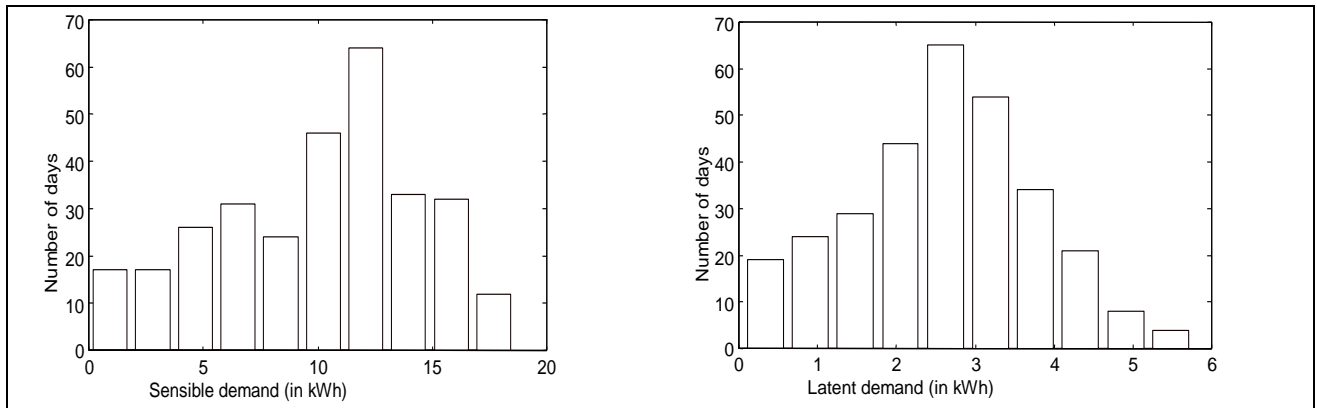

**Figure 11: Distribution of latent and sensible energy needs for the humid season.**

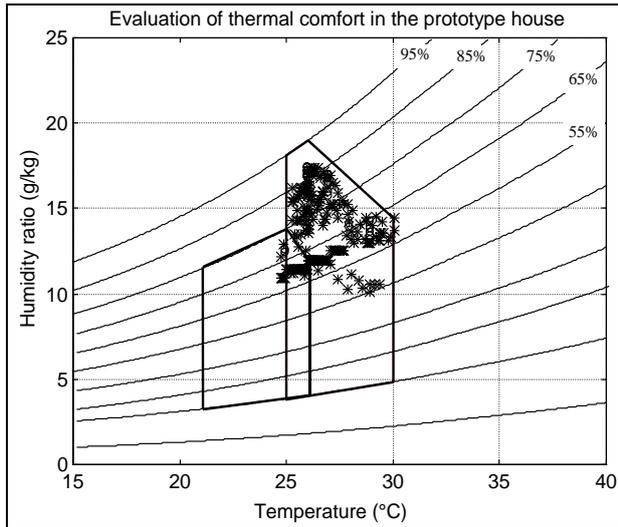 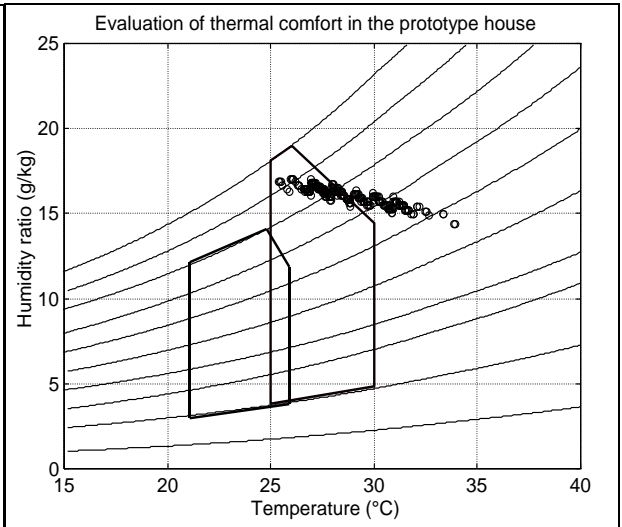

**Figure 12 : Thermal Comfort diagram for the sequence E**     **Figure 13 : For the sequences A and C**

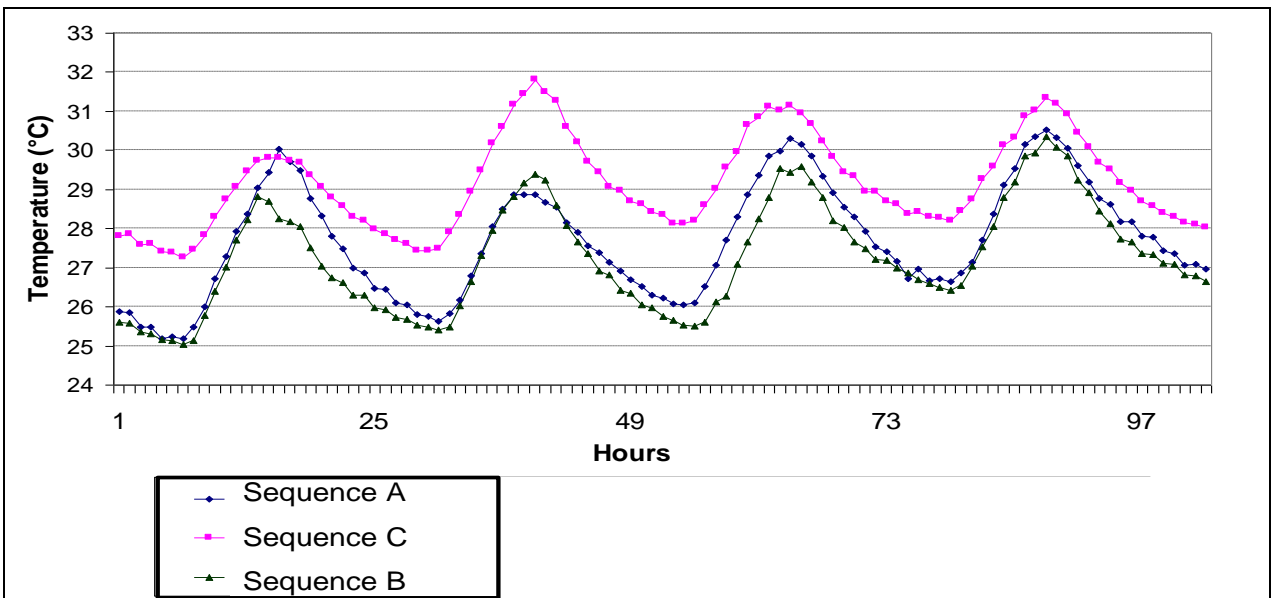

**Figure 14: Comparison of internal temperature in the house for the three sequences A, B, et C**